\documentclass[aps,prb,twocolumn,showpacs]{revtex4}

\usepackage{graphicx}% Include figure files
\usepackage{amsmath}
\usepackage{amssymb}

\renewcommand{\vec}[1]{\boldsymbol{#1}}

\begin{document}

\title{Doping and energy dependent microwave conductivity of kinetic
energy driven superconductors with extended impurities}

\author{Zhi Wang}
\affiliation{Department of Physics, Beijing Normal University,
Beijing 100875, China}

\author{Huaiming Guo}
\affiliation{Department of Physics, Capital Normal University,
Beijing 100037, China}

\author{Shiping Feng$^{*}$}
\affiliation{Department of Physics, Beijing Normal University,
Beijing 100875, China~~}

%\maketitle
%\date{\today}
\begin{abstract}
Within the framework of the kinetic energy driven superconducting
mechanism, the effect of the extended impurity scatterers on the
quasiparticle transport of cuprate superconductors in the
superconducting state is studied based on the nodal approximation of
the quasiparticle excitations and scattering processes. It is shown
that there is a cusplike shape of the energy dependent microwave
conductivity spectrum. At low temperatures, the microwave
conductivity increases linearly with increasing temperatures, and
reaches a maximum at intermediate temperature, then decreases with
increasing temperatures at high temperatures. In contrast with the
dome shape of the doping dependent superconducting gap parameter,
the minimum microwave conductivity occurs around the optimal doping,
and then increases in both underdoped and overdoped regimes.
\end{abstract}

%\bigskip

%\narrowtext
\maketitle

\section{Introduction}

After over 20 years extensive studies, it has become clear that
superconductivity in doped cuprates results when charge carriers
pair up into Cooper pairs \cite{tsuei} as in the conventional
superconductors \cite{bcs}, then these charge carrier Cooper pairs
condensation reveals the superconducting (SC) ground-state. However,
as a natural consequence of the unconventional SC mechanism that is
responsible for the high SC transition temperatures \cite{anderson},
the charge carrier Cooper pairs in cuprate superconductors have a
dominated d-wave symmetry \cite{tsuei,shen}. In particular, this
d-wave SC state remains one of the cornerstones of our understanding
of the physics in cuprate superconductors. However, in spite of the
unconventional SC mechanism, the angle-resolved photoemission
spectroscopy (ARPES) experimental results have unambiguously
established the Bogoliubov-quasiparticle nature of the sharp SC
quasiparticle peak in cuprate superconductors
\cite{matsui,campuzano1}, then the SC coherence of the low energy
quasiparticle peak is well described by the simple
Bardeen-Cooper-Schrieffer (BCS) formalism \cite{bcs} with the d-wave
SC gap function $\Delta({\bf k})=\Delta({\rm cos}k_{x}- {\rm
cos}k_{y})/2$. In this d-wave case, the characteristic feature is
the existence of four nodal points $[\pm\pi/2,\pm\pi/2]$ (in units
of inverse lattice constant) in the Brillouin zone
\cite{tsuei,shen}, where the SC gap function vanishes, then the most
physical properties of cuprate superconductors in the SC state are
controlled by the quasiparticle excitations around the nodes
\cite{turner}. In particular, the key signature of the nodal
quasiparticle transport appears in the microwave conductivity
$\sigma(\omega,T)$ \cite{turner}, which is essentially
electromagnetic absorption by the quasiparticles excited out of the
condensation (either thermally excited quasiparticles or excitations
created by the impurity scattering).

Understanding the role of impurities in cuprate superconductors has
taken many years of great effort \cite{palee,hirschfeld}. This
follows from that the physical properties of cuprate superconductors
in the SC state are extreme sensitivity to the impurity effect than
the conventional superconductors due to the finite angular-momentum
charge carrier Cooper pairing \cite{hirschfeld1}. Experimentally, By
virtue of systematic studies using the microwave conductivity
measurements, some essential features of the evolution of the
quasiparticle transport of cuprate superconductors with energy and
temperature in the SC state have been established
\cite{turner,bonn,lee,hosseini,harris}: (1) at low temperatures, the
experimental results show the existence of the very long-live
excitation deep in the SC state, as evidenced by the sharp cusplike
energy dependent microwave conductivity spectrum, where the width of
the sharp peak is nearly temperature independent, and main behaviors
of the microwave conductivity are governed by thermally excited
quasiparticles being scattered by impurities or other defects
\cite{turner,harris}; (2) at low energies, the temperature dependent
microwave conductivity increases linearly with increasing
temperatures at low temperatures, and reaches a maximum (a large
broad peak) around intermediate temperature, then decreases with
increasing temperatures at high temperatures
\cite{bonn,hosseini,harris}. In particular, this broad peak shifts
to higher temperatures as the energy is increased. Theoretically, an
agreement has emerged that the BCS formalism with the d-wave SC gap
function is useful in the phenomenological description of the
quasiparticle transport of cuprate superconductors in the SC state
\cite{palee,hirschfeld,quinlan,hettler,durst,duffy,nunner}, although
the SC pairing mechanism is beyond the conventional electron-phonon
mechanism. In this case, the microwave conductivity of cuprate
superconductors has been phenomenologically discussed in the zero
temperature and energy by including the contributions of the vertex
corrections \cite{durst}. Recently, these discussions have been
generalized to study the temperature and energy dependence of the
quasiparticle transport of cuprate superconductors in the SC state
\cite{nunner}. To the best of our knowledge, the microwave
conductivity of cuprate superconductors has not been treated
starting from a microscopic SC theory, and no explicit calculations
of the doping dependence of the microwave conductivity has been made
so far.

In this paper, we start from the kinetic energy driven SC mechanism
\cite{feng1}, and study the effect of the extended impurity
scatterers on the microwave conductivity of cuprate superconductors.
We evaluate explicitly the microwave conductivity of cuprate
superconductors within the nodal approximation of the quasiparticle
excitations and scattering processes, and qualitatively reproduced
some main features of the microwave conductivity measurements on
cuprate superconductors in the SC state
\cite{turner,bonn,lee,hosseini,harris}. It is shown that there is a
cusplike shape of the energy dependent microwave conductivity
spectrum. At low temperatures, the microwave conductivity increases
linearly with increasing temperatures, and reaches a maximum at
intermediate temperature, then decreases with increasing
temperatures at high temperatures. In contrast with the dome shape
of the doping dependent SC gap parameter, the minimum microwave
conductivity occurs around the optimal doping, and then increases in
both underdoped and overdoped regimes.

This paper is organized as follows. We present the basic formalism
in Sec. II, and then discuss the energy, temperature, and doping
dependence of the quasiparticle transport of cuprate superconductors
in the SC state in Sec. III, where we show that the quasiparticle
transport of cuprate superconductors in the SC state can be
qualitatively understood within the framework of the kinetic energy
driven SC mechanism by considering the effect of the extended
impurity scatterers.  Finally, we give a summary in Sec. IV.

\section{Formalism}

In cuprate superconductors, the single common feature is the
presence of the two-dimensional CuO$_{2}$ plane, and it is believed
that the unconventional physical properties of cuprate
superconductors is closely related to the doped CuO$_{2}$ planes
\cite{shen,kastner}. It has been argued that the $t$-$J$ model on a
square lattice captures the essential physics of the doped CuO$_{2}$
plane \cite{anderson,shen},
\begin{equation}
H=-t\sum_{i\hat{\eta}\sigma}\hat{C}^{\dagger}_{i\sigma}
\hat{C}_{i+\hat{\eta}\sigma}+\mu \sum_{i\sigma}
\hat{C}^{\dagger}_{i\sigma}\hat{C}_{i\sigma}
+J\sum_{i\hat{\eta}}{\bf S}_{i} \cdot {\bf S}_{i+\hat{\eta}},
\end{equation}
where $\hat{\eta}=\pm\hat{x},\pm\hat{y}$, ${\bf S}_{i}=
\hat{C}^{\dagger}_{i}{\vec\tau}\hat{C}_{i}/2$ is spin operator with
${\vec{\tau}}=(\tau_{1},\tau_{2},\tau_{3})$ as Pauli matrices, the
constrained electron operator $\hat{C}_{i\sigma}=C_{i\sigma}(1-
n_{i-\sigma})$ with $n_{i\sigma}=C^{\dagger}_{i\sigma}C_{i\sigma}$,
and $\mu$ is the chemical potential. In the constrained electron
operator, the operators $C^{\dagger}_{i\sigma}$ and $C_{i\sigma}$
are to be thought of as operating within the full Hilbert space,
while the constrained electron operator
$\hat{C}^{\dagger}_{i\sigma}$ ($\hat{C}_{i\sigma}$) does not create
(destroy) any doubly occupied sites, and therefore represents
physical creation (annihilation) operator acting in the restricted
Hilbert space without double electron occupancy
\cite{rice,anderson1}. The strong electron correlation in the
$t$-$J$ model manifests itself by the restriction of the motions of
the electrons in the restricted Hilbert space without double
electron occupancy \cite{anderson}, which can be treated properly in
analytical calculations within the charge-spin separation (CSS)
fermion-spin theory \cite{feng2}, where the constrained electron
operators are decoupled as $\hat{C}_{i\uparrow}=
h^{\dagger}_{i\uparrow}S^{-}_{i}$ and $\hat{C}_{i\downarrow}=
h^{\dagger}_{i\downarrow} S^{+}_{i}$, with the spinful fermion
operator $h_{i\sigma}=e^{-i\Phi_{i\sigma}} h_{i}$ describes the
charge degree of freedom together with some effects of spin
configuration rearrangements due to the presence of the doped hole
itself (dressed holon), while the spin operator $S_{i}$ describes
the spin degree of freedom (spin), then the motions of electrons are
restricted in the restricted Hilbert space without double electron
occupancy in analytical calculations. In particular, it has been
shown that under the decoupling scheme, this CSS fermion-spin
representation is a natural representation of the constrained
electron defined in the restricted Hilbert space without double
electron occupancy \cite{feng3}. Moreover, these dressed holon and
spin are gauge invariant \cite{feng2}, and in this sense, they are
real and can be interpreted as the physical excitations
\cite{laughlin}. In this CSS fermion-spin representation, the
low-energy behavior of the $t$-$J$ model (1) can be expressed as,
\begin{eqnarray}
H&=&t\sum_{i\hat{\eta}}(h^{\dagger}_{i+\hat{\eta}\uparrow}
h_{i\uparrow}S^{+}_{i}S^{-}_{i+\hat{\eta}}+
h^{\dagger}_{i+\hat{\eta}\downarrow}h_{i\downarrow}S^{-}_{i}
S^{+}_{i+\hat{\eta}})\nonumber \\
&-&\mu\sum_{i\sigma}h^{\dagger}_{i\sigma} h_{i\sigma}+J_{{\rm eff}}
\sum_{i\hat{\eta}}{\bf S}_{i}\cdot {\bf S}_{i+\hat{\eta}},
\end{eqnarray}
with $J_{{\rm eff}}=(1-\delta)^{2}J$, and $\delta=\langle
h^{\dagger}_{i\sigma}h_{i\sigma}\rangle=\langle h^{\dagger}_{i}
h_{i}\rangle$ is the hole doping concentration. As an important
consequence, the kinetic energy term in the $t$-$J$ model has been
transferred as the dressed holon-spin interaction, which reflects
that even the kinetic energy term in the $t$-$J$ Hamiltonian has
strong Coulombic contribution due to the restriction of the motions
of electrons in the restricted Hilbert space without double electron
occupancy.

Recently, we have developed a kinetic energy driven SC mechanism
\cite{feng1} based on the CSS fermion-spin theory \cite{feng2},
where the dressed holon-spin interaction from the kinetic energy
term in the $t$-$J$ model (2) induces the dressed holon pairing
state with the d-wave symmetry by exchanging spin excitations, then
the electron Cooper pairs originating from the dressed holon pairing
state are due to the charge-spin recombination, and their
condensation reveals the d-wave SC ground-state. Moreover, this
d-wave SC state is controlled by both SC gap function and
quasiparticle coherence, then the maximal SC transition temperature
occurs around the optimal doping, and decreases in both underdoped
and overdoped regimes. In particular, we have shown that this SC
state is the conventional BCS like with the d-wave symmetry
\cite{feng3,guo1}, so that the basic BCS formalism with the d-wave
SC gap function is still valid in quantitatively reproducing all
main low energy features of the ARPES experimental measurements on
cuprate superconductors, although the pairing mechanism is driven by
the kinetic energy by exchanging spin excitations, and other exotic
magnetic scattering \cite{dai,arai} is beyond the BCS formalism.
Following our previous discussions \cite{feng1,feng3,guo1}, the full
dressed holon Green's function in the SC state can be obtained in
the Nambu representation as,
\begin{eqnarray}
\tilde{g}({\bf k},\omega)&=&Z_{hF}{1\over\omega^{2}-E^{2}_{h{\bf k}}
} \left(
\begin{array}{cc}
{\omega+\bar{\xi}_{{\bf k}}} & {\bar{\Delta}_{hZ}({\bf k})} \\
{\bar{\Delta}_{hZ}({\bf k})} & {\omega-\bar{\xi}_{{\bf k}}}
\end{array}\right)\nonumber \\
&=&Z_{hF}{\omega\tau_{0}+\bar{\Delta}_{hZ} ({\bf k})
\tau_{1}+\bar{\xi}_{{\bf k}}\tau_{3}\over\omega^{2}- E^{2}_{h{\bf k}
}},
\end{eqnarray}
where $\tau_{0}$ is the unit matrix, the renormalized dressed holon
excitation spectrum $\bar{\xi}_{{\bf k}}=Z_{hF}\xi_{\bf k}$, with
the mean-field (MF) dressed holon excitation spectrum $\xi_{{\bf k}}
=Zt\chi\gamma_{{\bf k}}-\mu$, the spin correlation function $\chi=
\langle S_{i}^{+}S_{i+\hat{\eta}}^{-}\rangle$, $\gamma_{{\bf k}}=
(1/Z)\sum_{\hat{\eta}}e^{i{\bf k}\cdot \hat{\eta}}$, $Z$ is the
number of the nearest neighbor sites, the renormalized dressed holon
d-wave pair gap function $\bar{\Delta}_{hZ}({\bf k})=Z_{hF}
\bar{\Delta}_{h}({\bf k})$, where the effective dressed holon d-wave
pair gap function $\bar{\Delta}_{h}({\bf k})=\bar{\Delta}_{h}
\gamma^{(d)}_{{\bf k}}$ with $\gamma^{(d)}_{{\bf k}}=({\rm cos}
k_{x}-{\rm cos}k_{y})/2$, and the dressed holon quasiparticle
spectrum $E_{h{\bf k}}=\sqrt {\bar{\xi}^{2}_{{\bf k}}+
\mid\bar{\Delta}_{hZ}({\bf k})\mid^{2}}$, while the dressed holon
quasiparticle coherent weight $Z_{hF}$ and effective dressed holon
gap parameters $\bar{\Delta}_{h}$ are determined by the following
two equations \cite{feng1,feng3,guo1},
\begin{widetext}
\begin{subequations}
\begin{eqnarray}
1&=&{(Zt)^{2}\over N^{3}}\sum_{{\bf k,p,p'}}\gamma^{2}_{{\bf p+k}}
\gamma^{(d)}_{{\bf k-p'+p}}\gamma^{(d)}_{{\bf k}}{Z^{2}_{hF}\over
E_{h{\bf k}}}{B_{{\bf p}}B_{{\bf p'}}\over\omega_{{\bf p}}
\omega_{{\bf p'}}}\left({F^{(1)}_{1}({\bf k,p,p'})\over
(\omega_{{\bf p'}}- \omega_{{\bf p}})^{2}-E^{2}_{h{\bf k}}}-
{F^{(2)}_{1}({\bf k,p,p'}) \over (\omega_{{\bf p'}} + \omega_{{\bf
p}})^{2}-E^{2}_{h{\bf k}}}
\right ) ,\\
{1\over Z}_{hF} &=& 1+\left({Zt\over N}\right)^{2}\sum_{{\bf p,p'}}
\gamma^{2}_{{\bf p}+{\bf k}_{0}}Z_{hF}{B_{{\bf p}}B_{{\bf p'}}\over
4\omega_{{\bf p}}\omega_{{\bf p'}}}\left({F^{(1)}_{2}({\bf p,p'})
\over (\omega_{{\bf p}}-\omega_{{\bf p'}}-E_{h{\bf p-p'+k_{0}}}
)^{2}}+{F^{(2)}_{2}({\bf p,p'})\over (\omega_{{\bf p}}-\omega_{{\bf
p'}} +E_{h{\bf p-p'+k_{0}}})^{2}}\right . \nonumber \\
&+& \left . {F^{(3)}_{2}({\bf p,p'})\over (\omega_{{\bf p}}+
\omega_{{\bf p'}}-E_{h{\bf p-p'+k_{0}}})^{2}}+{F^{(4)}_{2} ({\bf p,
p'}) \over (\omega_{{\bf p}} + \omega_{{\bf p'}} + E_{h{\bf
p-p'+k_{0} }})^{2}} \right ) ,
\end{eqnarray}
\end{subequations}
\end{widetext}
respectively, where ${\bf k}_{0}=[\pi,0]$, $B_{{\bf p}}=\lambda[(2
\epsilon\chi^{z}+\chi)\gamma_{{\bf p}}-(2\chi^{z}+\epsilon\chi)]$,
$\lambda=2ZJ_{{\rm eff}}$, $\epsilon=1+2t\phi/J_{{\rm eff}}$, the
dressed holon's particle-hole parameter $\phi=\langle
h^{\dagger}_{i\sigma}h_{i+\hat{\eta}\sigma}\rangle$, the spin
correlation function $\chi^{z}=\langle S_{i}^{z}S_{i+\hat{\eta}}^{z}
\rangle$, $F^{(1)}_{1}({\bf k,p,p'})= (\omega_{{\bf p'}}-
\omega_{{\bf p}})[n_{B}(\omega_{{\bf p}})-n_{B}(\omega_{{\bf p'}})]
[1-2n_{F}(E_{h{\bf k}})]+E_{h{\bf k}}[n_{B} (\omega_{{\bf p'}})
n_{B}(-\omega_{{\bf p}})+n_{B} (\omega_{{\bf p}})
n_{B}(-\omega_{{\bf p'}})]$, $F^{(2)}_{1}({\bf k, p,p'})=
(\omega_{{\bf p'}}+\omega_{{\bf p}})[n_{B}(-\omega_{{\bf p'}} )-
n_{B}(\omega_{{\bf p}})][1-2n_{F} (E_{h{\bf k}})]+E_{h{\bf k}}[n_{B}
(\omega_{{\bf p'}})n_{B}(\omega_{{\bf p}})+n_{B}(-\omega_{{\bf p'}})
n_{B}(-\omega_{{\bf p}})]$, $F^{(1)}_{2}({\bf p,p'})=n_{F}(E_{h{\bf
p-p'+k_{0}}})[n_{B}(\omega_{{\bf p'}})-n_{B} (\omega_{{\bf p}})]-
n_{B}(\omega_{{\bf p}})n_{B}(-\omega_{{\bf p'}} )$, $F^{(2)}_{2}
({\bf p,p'})=n_{F}(E_{h{\bf p-p'+k_{0}}}) [n_{B} (\omega_{{\bf p}})
-n_{B}(\omega_{{\bf p'}})]-n_{B}(\omega_{{\bf p'}} )n_{B}
(-\omega_{{\bf p}})$, $F^{(3)}_{2}({\bf p,p'})=n_{F}(E_{h{\bf p-p'+
k_{0}}})[n_{B}(\omega_{{\bf p'}})-n_{B}(-\omega_{{\bf p}})]+ n_{B}
(\omega_{{\bf p}})n_{B}(\omega_{{\bf p'}})$, $F^{(4)}_{2}({\bf p,p'}
)=n_{F}(E_{h{\bf p-p'+k_{0}}})[n_{B} (-\omega_{{\bf p'}})-n_{B}
(\omega_{{\bf p}})]+n_{B}(-\omega_{{\bf p}})n_{B}(-\omega_{{\bf p'}}
)$, $n_{B}(\omega_{{\bf p}})$ and $n_{F}(E_{h{\bf k}})$ are the
boson and fermion distribution functions, respectively, and the MF
spin excitation spectrum,
\begin{eqnarray}
\omega^{2}_{{\bf p}}&=&\lambda^{2}[(A_{1}-\alpha\epsilon\chi^{z}
\gamma_{{\bf p}}-{1\over 2Z}\alpha\epsilon\chi) (1-\epsilon
\gamma_{{\bf p}})\nonumber \\
&+&{1\over 2}\epsilon(A_{2}-{1\over 2} \alpha\chi^{z}
-\alpha\chi\gamma_{{\bf p}})(\epsilon-\gamma_{{\bf p}})],
\end{eqnarray}
where $A_{1}=\alpha C^{z}+(1-\alpha)/(4Z)$, $A_{2}=\alpha C+
(1-\alpha)/(2Z)$, and the spin correlation functions $C=(1/Z^{2})
\sum_{\hat{\eta},\hat{\eta'}}\langle S_{i+\hat{\eta}}^{+}
S_{i+\hat{\eta'}}^{-}\rangle$ and $C^{z}=(1/Z^{2})
\sum_{\hat{\eta},\hat{\eta'}}\langle S_{i+\hat{\eta}}^{z}
S_{i+\hat{\eta'}}^{z}\rangle$. In order to satisfy the sum rule of
the correlation function $\langle S^{+}_{i}S^{-}_{i}\rangle=1/2$ in
the case without the antiferromagnetic long-range-order, an
important decoupling parameter $\alpha$ has been introduced in the
above MF calculation \cite{feng1,feng3,guo1}, which can be regarded
as the vertex correction. These two equations in Eqs. (4a) and (4b)
must be solved simultaneously with other self-consistent equations,
then all order parameters, decoupling parameter $\alpha$, and
chemical potential $\mu$ are determined by the self-consistent
calculation.

In the CSS fermion-spin theory \cite{feng2,feng3}, the electron
Green's function is a convolution of the spin Green's function and
dressed holon Green's function. Following our previous discussions
\cite{feng1,feng3,guo1}, we can obtain the electron diagonal and
off-diagonal Green's functions in the SC state as,
\begin{widetext}
\begin{subequations}
\begin{eqnarray}
G({\bf k},\omega)&=&{1\over N}\sum_{{\bf p}}Z_{hF}{B_{{\bf p}}\over
4\omega_{{\bf p}}}\left\{{\rm coth}[{1\over 2}\beta\omega_{{\bf p}}]
\left({U^{2}_{h{\bf p+k}}\over\omega+E_{h{\bf p+k}}-\omega_{{\bf p}}
}+{U^{2}_{h{\bf p+k}}\over\omega+E_{h{\bf p+k}}+\omega_{{\bf p}}}
\right . \right .\nonumber \\
&+& \left . {V^{2}_{h{\bf p+k}}\over\omega-E_{h{\bf p+k}}+
\omega_{{\bf p}}}+{V^{2}_{h{\bf p+k}}\over\omega-E_{h{\bf p+k}}
-\omega_{{\bf p}}}\right)+{\rm tanh}[{1\over 2}\beta E_{h{\bf p+k}}]
\left({U^{2}_{h{\bf p+k}}\over\omega+E_{h{\bf p+k}}+\omega_{{\bf p}}
}\right . \nonumber \\
&-&{U^{2}_{h{\bf p+k}}\over\omega+E_{h{\bf p+k}}-\omega_{{\bf p}}}
+\left . \left . {V^{2}_{h{\bf p+k}}\over\omega-E_{h{\bf p+k}}-
\omega_{{\bf p}}}-{V^{2}_{h{\bf p+k}}\over\omega-E_{h{\bf p+k}}
+\omega_{{\bf p}}} \right ) \right \} ,
\end{eqnarray}
\begin{eqnarray}
\Gamma^{\dagger}({\bf k},\omega)&=&{1\over N}\sum_{{\bf p}}Z_{hF}
{\bar{\Delta}_{hZ}({\bf p+k})\over 2E_{h{\bf p+k}}}{B_{{\bf p }}
\over 4\omega_{{\bf p}}}\left \{{\rm coth}[{1\over 2}\beta
\omega_{{\bf p}}]\left({1\over\omega-E_{h{\bf p+k}}-\omega_{{\bf p}}
}+{1\over \omega-E_{h{\bf p+k}}+\omega_{{\bf p}}}
\right .\right .\nonumber \\
&-& \left . {1\over \omega +E_{h{\bf p+k}}+\omega_{{\bf p}}}-
{1\over \omega +E_{h{\bf p+k}}-\omega_{{\bf p}}} \right )+{\rm
tanh}[{1\over 2}\beta E_{h{\bf p+k}}]\left ({1\over \omega -
E_{h{\bf p+k}} -\omega_{{\bf p}}} \right .\nonumber \\
&-&\left .\left . {1\over \omega-E_{h{\bf p+k}}+\omega_{{\bf p}}}
-{1\over \omega +E_{h{\bf p+k}}+\omega_{{\bf p}}}+ {1\over \omega
+E_{h{\bf p+k}}-\omega_{{\bf p}}} \right )\right \},
\end{eqnarray}
\end{subequations}
\end{widetext}
respectively, where the dressed holon quasiparticle coherence
factors $U^{2}_{h{\bf k}}=(1+{\bar{\xi}_{{\bf k}}/E_{h{\bf k}}})/2$
and $V^{2}_{h{\bf k}}=(1-{\bar{\xi}_{{\bf k}}/E_{h{\bf k}}})/2$.
These convolutions of the spin Green's function and dressed holon
diagonal and off-diagonal Green's functions reflect the charge-spin
recombination \cite{anderson2}. Since the spins center around the
$[\pi,\pi]$ point in the MF level \cite{feng1,feng3,guo1}, then the
main contributions for the spins comes from the $[\pi,\pi]$ point.
In this case, the electron diagonal and off-diagonal Green's
functions in Eqs. (6a) and (6b) can be approximately reduced as the
BCS formalism with the d-wave SC gap function in terms of
$\omega_{{\bf p}=[\pi,\pi]}\sim 0$ and the equation
\cite{feng1,feng2} $1/2=\langle S_{i}^{+}S_{i}^{-}\rangle=(1/N)
\sum_{{\bf p}}B_{{\bf p}}{\rm coth}(\beta \omega_{{\bf p}}/2)
/(2\omega_{{\bf p}})$,
\begin{subequations}
\begin{eqnarray}
G({\bf k},\omega)&\approx&Z_{F}\left ({U^{2}_{{\bf k}}\over\omega-
E_{{\bf k}}}+{V^{2}_{{\bf k}}\over \omega+E_{{\bf k}}}\right ),\\
\Gamma^{\dagger}({\bf k},\omega)&\approx& Z_{F}{\bar{\Delta}_{Z}
({\bf k})\over 2E_{{\bf k}}}\left ({1\over \omega-E_{{\bf k}}}+
{1\over \omega+E_{{\bf k}}}\right ),
\end{eqnarray}
\end{subequations}
where the electron quasiparticle coherent weight $Z_{F}=Z_{hF}/2$,
the electron quasiparticle coherence factors $U^{2}_{{\bf k}}
\approx V^{2}_{h{\bf k+k_{A}}}=(1+{\bar{\varepsilon}_{{\bf k}}/
E_{{\bf k}}})/2$ and $V^{2}_{{\bf k}}\approx U^{2}_{h{\bf k+k_{A}}}
=(1-{\bar{\varepsilon}_{{\bf k}}/E_{{\bf k}}})/2$, with
${\bar{\varepsilon}_{{\bf k}}=Z_{F}\varepsilon}_{{\bf k}}$,
$\varepsilon_{{\bf k}}=Zt\chi\gamma_{{\bf k}}+\mu$, and ${\bf k_{A}}
=[\pi,\pi]$, $\bar{\Delta}_{Z}({\bf k})=\bar{\Delta}_{hZ}({\bf k})
/2$, and electron quasiparticle spectrum $E_{{\bf k}}\approx
E_{h{\bf k+k_{A}}}=\sqrt{\bar{\varepsilon}^{2}_{{\bf k}}+
\mid\bar{\Delta}_{Z}({\bf k })\mid^{2}}$, i.e., the hole-like
dressed holon quasiparticle coherence factors $V_{h{\bf k}}$ and
$U_{h{\bf k}}$ and hole-like dressed holon quasiparticle spectrum
$E_{h{\bf k}}$ have been transferred into the electron quasiparticle
coherence factors $U_{{\bf k}}$ and $V_{{\bf k}}$ and electron
quasiparticle spectrum $E_{{\bf k}}$, respectively, by the
convolutions of the spin Green's function and dressed holon Green's
functions due to the charge-spin recombination. This means that
within the kinetic energy driven SC mechanism, the dressed holon
pairs condense with the d-wave symmetry in a wide range of the hole
doping concentration, then the electron Cooper pairs originating
from the dressed holon pairing state are due to the charge-spin
recombination, and their condensation automatically gives the
electron quasiparticle character. For the convenience in the
following discussions, these electron Green's functions in Eq. (7)
in the SC state can be expressed in the Nambu representation as,
\begin{eqnarray}
\tilde{G}({\bf k},\omega)&=&Z_{F}{1\over\omega^{2}-E^{2}_{{\bf k}}}
\left(
\begin{array}{cc}
{\omega+\bar{\varepsilon}_{{\bf k}}}&{\bar{\Delta}_{Z}({\bf k})}\\
{\bar{\Delta}_{Z}({\bf k})} & {\omega-\bar{\varepsilon}_{{\bf k}}}
\end{array} \right)\nonumber\\
&=&Z_{F}{\omega\tau_{0}+ \bar{\Delta}_{Z} ({\bf k})\tau_{1}+
\bar{\varepsilon}_{{\bf k}}\tau_{3}\over \omega^{2}-E_{{\bf k}
}^{2}}.
\end{eqnarray}

With the helps of this BCS formalism under kinetic energy driven SC
mechanism, now we can discuss the effect of the extended impurity
scatterers on the quasiparticle transport in cuprate
superconductors. In the presence of impurities, the unperturbed
electron Green's function in Eq. (8) is dressed via the impurity
scattering \cite{durst,nunner},
\begin{eqnarray}
\tilde{G}_{I}({\bf k},\omega)^{-1}=\tilde{G}({\bf k},\omega)^{-1}
-\tilde{\Sigma}({\bf k},\omega),
\end{eqnarray}
with the self-energy function $\tilde{\Sigma}({\bf k},\omega)=
\sum_{\alpha}\Sigma_{\alpha}({\bf k} ,\omega)\tau_{\alpha}$. It has
been shown that all but the scalar component of the self-energy
function can be neglected or absorbed into $\bar{\Delta}_{Z}({\bf k}
)$ \cite{durst,nunner}. In this case, the dressed electron Green's
function in Eq. (9) can be explicitly rewritten as,
\begin{widetext}
\begin{eqnarray}
\tilde{G}_{I}({\bf k},\omega)=Z_{F}{[\omega-\Sigma_{0}({\bf k},
\omega)]\tau_{0}+\bar{\Delta}_{Z}({\bf k})\tau_{1}+
[\bar{\varepsilon}_{{\bf k}}+\Sigma_{3}({\bf k},\omega)]\tau_{3}
\over [\omega-\Sigma_{0}({\bf k},\omega)]^{2}
-\bar{\varepsilon}^{2}_{{\bf k}}-\bar{\Delta}^{2}_{Z}({\bf k})}.
\end{eqnarray}
\end{widetext}
Based on the phenomenological d-wave BCS-type electron Green's
function \cite{durst}, the energy and temperature dependence of the
microwave conductivity of cuprate superconductors has been fitted
\cite{nunner}, where the electron self-energy functions
$\Sigma_{0}({\bf k},\omega)$ and $\Sigma_{3}({\bf k},\omega)$ have
been treated within the framework of the T-matrix approximation.
Following their discussions \cite{durst,nunner}, the electron
self-energy function $\tilde{\Sigma}({\bf k},\omega)$ can be
obtained approximately as,
\begin{eqnarray}
\tilde{\Sigma}({\bf k},\omega)=\rho_{i}\tilde{T}_{{\bf k}{\bf k}}
(\omega),
\end{eqnarray}
with $\rho_{i}$ is the impurity concentration, and $\tilde{T}_{{\bf
k}{\bf k}}(\omega)$ is the diagonal element of the T-matrix,
\begin{eqnarray}
\tilde{T}_{{\bf k}{\bf k}'}(\omega)=V_{{\bf k}{\bf k}'}\tau_{3}+
\sum_{{\bf k}''}V_{{\bf k}{\bf k}''}\tau_{3}\tilde{G}_{I}({\bf k}'',
\omega)\tilde{T}_{{\bf k}'' {\bf k}'}(\omega),
\end{eqnarray}
where $V_{{\bf k}{\bf k}'}$ is the impurity scattering potential. As
mentioned in Sec. I, there is no gap to the quasiparticle
excitations at the four nodes for the d-wave SC state of cuprate
superconductors, therefore the quasiparticles are generated only
around these four nodes. It has been shown \cite{durst} that this
characteristic feature is very useful when considering the impurity
scattering, since the initial and final momenta of a scattering
event must always be approximately equal to the ${\bf k} $-space
location of one of the four nodes in the zero temperature and zero
energy, while the impurity scattering potential $V_{{\bf k}{\bf k}'}
$ varies slowly over the area of a node. In this case, a general
scattering potential $V_{{\bf k}{\bf k}'}$ need only be evaluated in
three possible case: the intranode impurity scattering $V_{{\bf k}
{\bf k}'}=V_{1}$ (${\bf k}$ and ${\bf k}'$ at the same node), the
adjacent-node impurity scattering $V_{{\bf k}{\bf k}'}=V_{2}$ (${\bf
k}$ and ${\bf k}'$ at the adjacent nodes), and the opposite-node
impurity scattering $V_{{\bf k}{\bf k}'}=V_{3}$ (${\bf k}$ and ${\bf
k}'$ at the opposite nodes), then the impurity scattering potential
$V_{{\bf k}{\bf k}'}$ in the T-matrix can be effectively reduced as
\cite{durst},
\begin{eqnarray}
V_{{\bf k}{\bf k}'}\rightarrow\underline{V}=\left(
\begin{array}{cccc}
V_{1} & V_{2} & V_{3} & V_{2} \\
V_{2} & V_{1} & V_{2} & V_{3} \\
V_{3} & V_{2} & V_{1} & V_{2} \\
V_{2} & V_{3} & V_{2} & V_{1}
\end{array} \right) \,.
\end{eqnarray}
At the zero temperature and zero energy, these nodes reduce to
points. In this case, this nodal approximation for the impurity
potential can reproduce any impurity potential. However, at finite
temperatures and energies, there is a limitation on the forward
scattering character of the impurity potential because this nodal
approximation assumes the Brillouin zone quadrant around a
particular node \cite{durst}. It has been shown \cite{nunner} that
although the strict forward scattering limit can therefore not be
reached at finite temperatures and energies, this nodal
approximation is still appropriate to treat the intermediate range
scatters. Therefore in the following discussions, we employ the
simplified impurity scattering potential in Eq. (13) to study the
impurity scattering effect on the quasiparticle transport of cuprate
superconductors. Substituting Eq. (13) into Eq. (12), the T-matrix
can be obtained as a $4\times4$-matrix around the nodal points,
\begin{eqnarray}
\tilde{T}_{jj'}(\omega)=V_{jj'}\tau_{3}+ \tilde{I}_{G}(\omega)
\tau_{3}\sum_{j''}V_{jj''} \tilde{T}_{j''j'}(\omega),
\end{eqnarray}
where $\tilde{I}_{G}(\omega)$ is the integral of the electron
Green's function, and can be obtained as,
\begin{eqnarray}
\tilde{I}_{G}(\omega)={1\over N}\sum_{{\bf k}}\tilde{G}_{I}({\bf k},
\omega)\approx\tilde{G}_{I0}(\omega)\tau_{0}+
\tilde{G}_{I3}(\omega)\tau_{3},
\end{eqnarray}
with $\tilde{G}_{I0}(\omega)$ and $\tilde{G}_{I3}(\omega)$ are given
by,
\begin{subequations}
\begin{eqnarray}
\tilde{G}_{I0}(\omega)&=& {1\over N}\sum_{{\bf k}}Z_{F}{\omega
-\Sigma_{0}(\omega)\over \omega^{2}-E^{2}_{{\bf k}}},\\
\tilde{G}_{I3}(\omega)&=&{1\over N}\sum_{{\bf k}}Z_{F}
{\bar{\varepsilon}_{{\bf k}}+\Sigma_{3}(\omega)\over \omega^{2}
-E^{2}_{{\bf k}}},
\end{eqnarray}
\end{subequations}
and the self-energy functions $\Sigma_{0}(\omega)$ and
$\Sigma_{0}(\omega)$ are evaluated as,
\begin{widetext}
\begin{subequations}
\begin{eqnarray}
\Sigma_{0}(\omega)&=&{\rho_{i}\over 4}\left({2\tilde{G}_{I0}(\omega)
V_{13}^{2}\over [1-\tilde{G}_{I3}(\omega)V_{13}]^{2}-
[\tilde{G}_{I0}(\omega)V_{13}]^{2}}+{\tilde{G}_{I0}(\omega)
(V^{-}_{123})^{2}\over [1-\tilde{G}_{I3}(\omega)V^{-}_{123}]^{2}-
[\tilde{G}_{I0}(\omega)V^{-}_{123}]^{2}}\right . \nonumber \\
&+&\left . {\tilde{G}_{I0}(\omega)(V^{+}_{123})^{2}\over [1-
\tilde{G}_{I3}(\omega)V^{+}_{123}]^{2}-[\tilde{G}_{I0}(\omega)
V^{+}_{123}]^{2}}\right ), \\
\Sigma_{3}(\omega)&=&{\rho_{i}\over 4}\left ( {2V_{13}[1-
\tilde{G}_{I3}(\omega) V_{13}]\over [1-
\tilde{G}_{I3}(\omega)V_{13}]^{2}- [\tilde{G}_{I0}(\omega)
V_{13}]^{2}}+{V^{-}_{123}[1-\tilde{G}_{I3} (\omega)V^{-}_{123}]\over
[1-\tilde{G}_{I3}(\omega)V^{-}_{123}]^{2}-
[\tilde{G}_{I0}(\omega)V^{-}_{123}]^{2}}\right . \nonumber\\
&+& \left . {V^{+}_{123}[1-\tilde{G}_{I0}(\omega)V^{+}_{123}]\over
[1-\tilde{G}_{I3}(\omega)V^{+}_{123}]^{2}- [\tilde{G}_{I0}(\omega)
V^{+}_{123}]^{2}} \right ),
\end{eqnarray}
\end{subequations}
\end{widetext}
where $V_{13}=V_{1}-V_{3}$, $V^{-}_{123}=V_{1}-2V_{2}+V_{3}$, and
$V^{+}_{123}=V_{1}+2V_{2}+V_{3}$.

In the framework of the linear response theory, the microwave
conductivity of cuprate superconductors can be calculated by means
of the Kubo formula as \cite{mahan},
\begin{eqnarray}
\sigma(\omega,T)=-{{\rm Im}\Pi(\omega,T)\over\omega},
\end{eqnarray}
with $\Pi(\omega,T)$ is the electron current-current correlation
function in the SC state, and can be expressed as,
\begin{eqnarray}
\Pi(\tau-\tau')=-<T_{\tau}{\bf J}(\tau)\cdot{\bf J}(\tau')>.
\end{eqnarray}
In the CSS fermion-spin representation \cite{feng2,feng3}, the
electron polarization operator can be evaluated as,
\begin{eqnarray}
{\bf P}=\sum_{i}{\bf R}_{i}\hat{n}_{i}=\sum_{i,\sigma}{\bf R}_{i}
\hat{C}^{\dag}_{i,\sigma}\hat{C}_{i,\sigma}={1\over 2}
\sum_{i,\sigma}{\bf R}_{i}h_{i,\sigma}h^{\dag}_{i,\sigma},~~~
\end{eqnarray}
then within the $t$-$J$ model (2), the current density of electrons
is obtained by the time derivation of this polarization operator
using the Heisenberg's equation of motion as,
\begin{widetext}
\begin{eqnarray}
{\bf J}=ie[H,{\bf P}]=i{1\over 2}et\sum_{i\hat{\eta}}\hat{\eta}
(h^{\dagger}_{i+\hat{\eta}\uparrow} h_{i\uparrow}S_{i}^{+}
S^{-}_{i+\hat{\eta}}+ h^{\dagger}_{i+\hat{\eta}\downarrow}
h_{i\downarrow}S^{\dagger}_{i}S^{-}_{i+\eta})=-i{1\over 2}et
\sum_{i\eta\sigma}\hat{\eta} \hat{C}^{\dagger}_{i\sigma}
\hat{C}_{i+\hat{\eta}\sigma}\approx {ev_{f}\over \sqrt{2}}
\sum_{{\bf k}\sigma}{\bf k} \hat{C}^{\dagger}_{{\bf k}\sigma}
\hat{C}_{{\bf k}\sigma},
\end{eqnarray}
\end{widetext}
with $v_{f}=\sqrt{2}t$ is the electron velocity at the nodal points.
According to this current density (21), the current-current
correlation function in Eq. (19) can be obtained as,
\begin{eqnarray}
\Pi(i\omega_{n},T)&=&{e^{2}v_{f}^{2}\over 2}{1\over N}\sum_{{\bf k}}
{1\over\beta}\sum_{i\omega_{n}'}{\rm Tr}[\tilde{G}({\bf k},
i\omega_{n}')\nonumber\\
&\times& \tilde{G}({\bf k},i\omega_{n}'+ i\omega_{n})
\tilde{\Gamma}({\bf k},i\omega_{n}',i\omega_{n})],
\end{eqnarray}
with $\tilde{\Gamma}({\bf k},\omega',\omega)$ is the vertex
function, and can be evaluated explicitly as the sum of ladder
diagrams,
\begin{eqnarray}
\Pi(i\omega_{n})=e^{2}v^{2}_{f}{1\over\beta}\sum_{i\omega_{n}'}
J(i\omega_{n},i\omega_{n}'),
\end{eqnarray}
where the kernel function $J(\omega,\omega')$ is expressed as,
\begin{widetext}
\begin{eqnarray}
J(\omega,\omega')={I^{(0)}_{0}+L_{A}[I^{(0)}_{0}I_{3}^{(3)}
+I_{0}^{(3)}I_{3}^{(0)}]\over [1-(L_{A}I_{0}^{(0)}+L_{B}
I^{(0)}_{3})][1-(L_{A}I_{3}^{(3)}+L_{B}I^{(3)}_{0})]-[L_{A}
I_{0}^{(3)}+L_{B}I^{(3)}_{3}][L_{A}I_{3}^{(0)}+L_{B}I^{(0)}_{0}]},
\end{eqnarray}
with the functions,
\begin{subequations}
\begin{eqnarray}
L_{A}(\omega,\omega')&=&[T_{11}^{(0)}(\omega)T_{11}^{(0)}(\omega+
\omega')+T_{11}^{(3)}(\omega)T_{11}^{(3)}(\omega+\omega') -
T_{13}^{(0)}(\omega)T_{13}^{(0)}(\omega+\omega')
-T_{13}^{(3)}(\omega)T_{13}^{(3)}(\omega+\omega')],\\
L_{B}(\omega,\omega')&=&[T_{11}^{(0)}(\omega)T_{11}^{(3)}(\omega+
\omega')+T_{11}^{(3)}(\omega)T_{11}^{(0)}(\omega+\omega')
-T_{13}^{(0)}(\omega)T_{13}^{(3)}(\omega+\omega')
-T_{13}^{(3)}(\omega)T_{13}^{(0)}(\omega+\omega')],
\end{eqnarray}
\end{subequations}
while the functions $I^{(0)}_{0}(\omega,\omega')$ and
$I^{(0)}_{3}(\omega,\omega')$ are give by,
\begin{subequations}
\begin{eqnarray}
I^{(0)}_{0}(\omega,\omega')\tau_{0}&+&I^{(3)}_{0}(\omega,\omega')
\tau_{3} ={1\over N}\sum_{{\bf k}}\tilde{G}_{I}({\bf k},\omega)
\tilde{G}_{I}({\bf k},\omega+\omega'), \\
I^{(0)}_{3}(\omega,\omega')\tau_{0}&+&I^{(3)}_{3}(\omega,\omega')
\tau_{3}={1\over N}\sum_{{\bf k}}\tilde{G}_{I}({\bf k},\omega)
\tilde{\tau}_{3}\tilde{G}_{I}({\bf k},\omega+\omega').
\end{eqnarray}
\end{subequations}
Substituting Eq. (23) into Eq. (18), the microwave conductivity of
cuprate superconductors is obtained explicitly as,
\begin{eqnarray}
\sigma(\omega,T)&=&e^{2}v^{2}_{f}\int^{\infty}_{-\infty}{d\omega'
\over2\pi}{n_{F}(\omega')-n_{F}(\omega' + \omega) \over \omega'}
[{\rm Re}J(\omega'-i\delta,\omega'+\omega+i\delta)-{\rm Re}
J(\omega'+i\delta,\omega'+\omega+i\delta)].
\end{eqnarray}
\end{widetext}
We emphasize that based on the nodal approximation of the
quasiparticle excitations and scattering processes
\cite{durst,nunner}, this microwave conductivity of cuprate
superconductors in Eq. (27) is obtained within the kinetic energy
driven SC mechanism, although its expression is similar to that
obtained within the phenomenological BCS formalism with the d-wave
SC gap function \cite{durst,nunner}.

\section{Microwave conductivity of cuprate superconductors}

In cuprate superconductors, although the values of $J$ and $t$ is
believed to vary somewhat from compound to compound, however, as a
qualitative discussion, the commonly used parameters in this paper
are chosen as $t/J=2.5$, with an reasonably estimative value of
$J\sim 1000$K. We are now ready to discuss the doping, energy, and
temperature dependence of the quasiparticle transport of cuprate
superconductors in the SC state with extended impurities. We have
performed a calculation for the microwave conductivity
$\sigma(\omega,T)$ in Eq. (27), and the results of
$\sigma(\omega,T)$ as a function of energy with temperature
$T=0.002J=2$K (solid line), $T=0.004J=4$K (dashed line),
$T=0.008J=8$K (dash-dotted line), and $T=0.01J=10$K (dotted line)
under the slightly strong impurity scattering potential with
$V_{1}=58J$, $V_{2}=49.32J$, and $V_{3}=40.6J$ at the impurity
concentration $\rho=0.000014$ for the doping concentration
$\delta=0.15$ are plotted in Fig. 1 in comparison with the
corresponding experimental results of cuprate superconductors in the
SC state \cite{harris} (inset). Obviously, the energy evolution of
the microwave conductivity of cuprate superconductor \cite{harris}
is qualitatively reproduced. In particular, a low temperature
cusplike shape of the microwave conductivity is obtained for cuprate
superconductors in the presence of the impurity scattering. At the
low energy regime ($\omega<0.0002J$), this low temperature microwave
conductivity $\sigma(\omega,T)$ rises rapidly from the universal
zero-temperature limit to a much larger microwave conductivity.
However, this low temperature microwave conductivity
$\sigma(\omega,T)$ becomes smaller and varies from weakly energy
dependence at the intermediate energy regime
($0.0002J<\omega<0.0008J$), to the almost energy independence at the
high energy regime ($\omega>0.0008J$).

\begin{figure}
\includegraphics[scale=0.55]{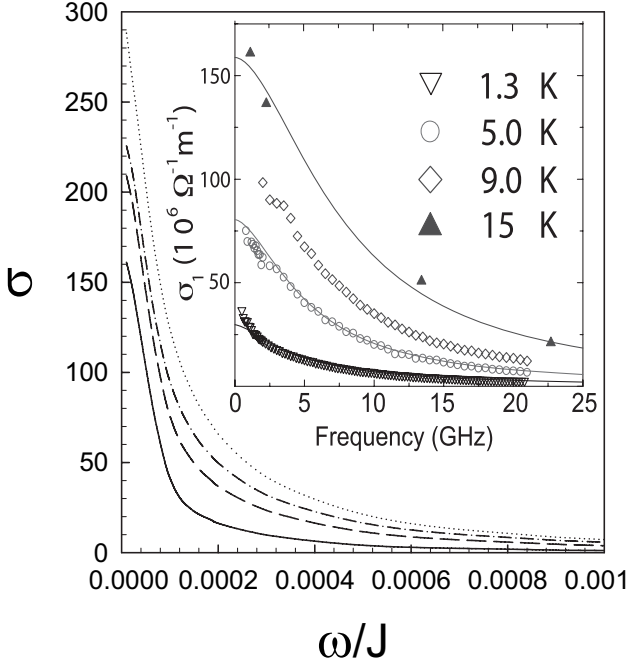}
\caption{The microwave conductivity as a function of energy with
$T=0.002J=2$K (solid line), $T=0.004J=4$K (dashed line), $T=0.008J
=8$K (dash-dotted line), and $T=0.01J=10$K (dotted line) at
$\rho=0.000014$ for $t/J=2.5$, $V_{1}=58J$, $V_{2}=49.32J$, and
$V_{3}=40.6J$ in $\delta=0.15$. Inset: the corresponding
experimental result of cuprate superconductors in the SC state taken
from Ref. [14].}
\end{figure}

\begin{figure}
\includegraphics[scale=0.50]{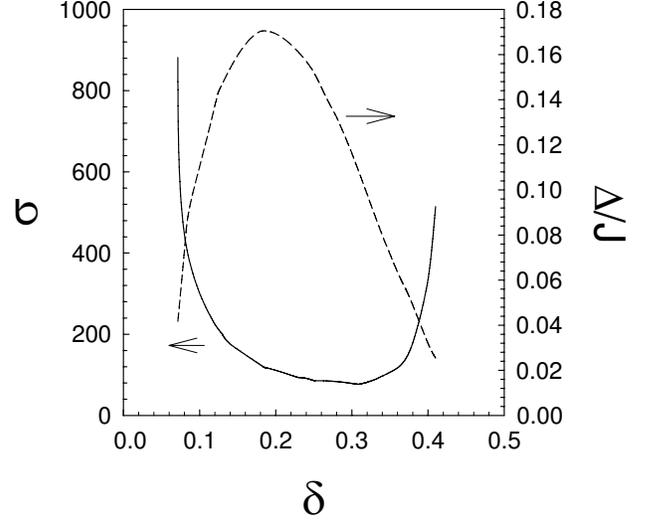}
\caption{The microwave conductivity as a function of doping with
$T=0.002J=$K and $\omega=0.000087J\approx 1.81$GHz at
$\rho=0.000014$ for $t/J=2.5$, $V_{1}=58J$, $V_{2}=49.32J$, and
$V_{3}=40.6J$. The dashed line is the corresponding result of the
superconducting gap parameter.}
\end{figure}

For a better understanding of the physical properties of the
microwave conductivity $\sigma(\omega,T)$ in cuprate
superconductors, we have studied the doping evolution of the
microwave conductivity, and the result of $\sigma(\omega,T)$ as a
function of doping with temperature $T=0.002J=2$K and energy
$\omega=0.000087J\approx 1.81$GHz under the slightly strong impurity
scattering potential with $V_{1}=58J$, $V_{2}=49.32J$, and
$V_{3}=40.6J$ at the impurity concentration $\rho=0.000014$ is
plotted in Fig. 2 (solid line). For comparison, the corresponding
result of the SC gap parameter of cuprate superconductors is also
shown in the same figure (dashed line). Our result shows that in
contrast to the dome shape of the doping dependent SC gap parameter
\cite{huefner,feng1,feng3,guo1}, the microwave conductivity
$\sigma(\omega,T)$ decreases with increasing doping in the
underdoped regime, and reaches a minimum in the optimal doping, then
increases in the overdoped regime. This doping dependent behavior of
the low energy microwave conductivity $\sigma(\omega,T)$ at low
temperatures is also qualitatively consistent with the universal
microwave conductivity limit \cite{palee} $\sigma\propto 1/\Delta$
at low energy as temperature $T\rightarrow 0$, if this SC gap
parameter $\Delta$ in the phenomenological BCS formalism
\cite{palee} has the similar dome shape doping dependence.

\begin{figure}
\includegraphics[scale=0.55]{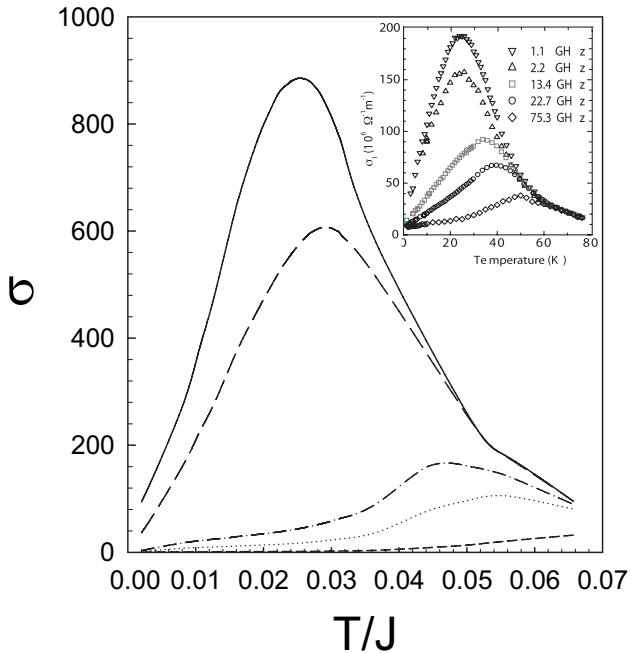}
\caption{The microwave conductivity as a function of temperature
with energy $\omega\approx 1.14$GHz (solid line), $\omega\approx
2.28$GHz (long dashed line), $\omega\approx 13.4$GHz (dash-dotted
line), $\omega\approx 22.8$GHz (dotted line), and $\omega\approx
75.3$GHz (short dashed line) at $\rho=0.000014$ for $t/J=2.5$,
$V_{1}=58J$, $V_{2}=49.32J$, and $V_{3}=40.6J$ in $\delta=0.15$.
Inset: the corresponding experimental result of cuprate
superconductors in the SC state taken from Ref. [14].}
\end{figure}

In the above discussions, we mainly study the effect of the extended
impurity scatterers on the quasiparticle transport at low
temperatures ($T\ll T_{c}$) and low energies ($\omega\ll\Delta$).
Now we discuss the temperature dependence of the quasiparticle
transport of cuprate superconductors, where $T$ may approach to
$T_{c}$ from low temperature side. In this case, it has been shown
\cite{nunner} that the inelastic scattering process should be
considered at higher temperatures, such as the
quasiparticle-quasiparticle scattering. This is followed from the
fact that the inelastic quasiparticle-quasiparticle scattering
process is suppressed at low temperatures due to the large SC gap
parameter in the quasiparticle excitation spectrum. However, the
contribution from this inelastic quasiparticle-quasiparticle
scattering process is increased rapidly when $T$ approaches to
$T_{c}$ from low temperature side, since there is a small SC gap
parameter near $T_{c}$. In particular, it has been pointed out
\cite{walker} that the contribution from the
quasiparticle-quasiparticle scattering process to the transport
lifetime is exponentially suppressed at low temperatures, and
therefore the effect of this inelastic quasiparticle-quasiparticle
scattering can be considered by adding the inverse transport
lifetime \cite{duffy} $\tau^{-1}_{{\rm inel}}(T)$ to the imaginary
part of the self-energy function $\Sigma_{0}(\omega)$ in Eq. (17a),
then the total self-energy function $\Sigma^{{\rm tot}}_{0}(\omega)$
can be expressed as \cite{walker,nunner},
\begin{eqnarray}
\Sigma^{{\rm tot}}_{0}(\omega)=\Sigma_{0}(\omega)-i[2\tau_{{\rm
inel}}(T)]^{-1},
\end{eqnarray}
with $\tau_{{\rm inel}}(T)$ has been chosen as $[2\tau_{{\rm inel}}
(T)]^{-1}=91.35(T-0.005)^{4}J$. Using this total self-energy
function $\Sigma^{{\rm tot}}_{0} (\omega)$ to replace
$\Sigma_{0}(\omega)$ in Eq. (27), we have performed a calculation
for the microwave conductivity $\sigma(\omega,T)$, and the results
of $\sigma(\omega,T)$ as a function of temperature $T$ with energy
$\omega=0.0000547J\approx 1.14$GHz (solid line), $\omega=0.0001094J
\approx 2.28$GHz (long dashed line), $\omega=0.0006564J\approx
13.4$GHz (dash-dotted line), $\omega=0.001094J\approx 22.8$GHz
(dotted line), and $\omega=0.0036102J\approx 75.3$GHz (short dashed
line) under the slightly strong impurity scattering potential with
$V_{1}=58J$, $V_{2}=49.32J$, and $V_{3}=40.6J$ at the impurity
concentration $\rho=0.000014$ for the doping concentration
$\delta=0.15$ are plotted in Fig. 3 in comparison with the
corresponding experimental results of cuprate superconductors in the
SC state \cite{harris} (inset). Our results show that the
temperature dependent microwave conductivity $\sigma(\omega,T)$
increases rapidly with increasing temperatures to a large broad peak
around temperature $T\approx 0.25J\approx 25$K for energy
$\omega=0.0000547J\approx 1.14$GHz, and then falls roughly linearly.
However, this broad peak in the microwave conductivity spectrum
resulting from the temperature dependent impurity scattering rate in
Eq. (28) is energy dependent, and moves to higher temperatures with
increasing energies, in qualitative agreement with the experimental
data \cite{harris}.

Within the framework of the kinetic energy driven d-wave cuprate
superconductivity \cite{feng1}, our present results of the energy
and temperature dependence of the microwave conductivity by
considering the effect of the extended impurity scatterers are
qualitatively similar to the earlier attempts to fit the
experimental data by using a phenomenological d-wave BCS formalism
\cite{palee,hirschfeld,quinlan,hettler,durst,duffy,nunner}.
Establishing this agreement is important to confirming the nature of
the SC phase of cuprate superconductors as the d-wave BCS-like SC
state within the kinetic energy driven SC mechanism. It has been
shown \cite{turner,bonn,lee,hosseini,harris} that there are some
subtle differences for different families of cuprate
superconductors, and these subtle differences may be induced by the
other effects except the impurity scattering. However, we in this
paper are primarily interested in exploring the general notion of
the effect of the extended impurity scatterers on the kinetic energy
driven cuprate superconductors in the SC state. The qualitative
agreement between the present theoretical results and experimental
data also show that the presence of impurities has a crucial effect
on the microwave conductivity of cuprate superconductors.

\section{Summary}

In conclusion we have shown very clearly in this paper that if the
effect of the extended impurity scatterers is taken into account in
the framework of the kinetic energy driven d-wave superconductivity,
the microwave conductivity of the $t$-$J$ model calculated based on
the nodal approximation of the quasiparticle excitations and
scattering processes per se can correctly reproduce some main
features found in the microwave conductivity measurements on cuprate
superconductor in the SC state
\cite{turner,bonn,lee,hosseini,harris}, including the energy and
temperature dependence of the microwave conductivity spectrum. The
theory also predicts a V-shaped doping dependent microwave
conductivity, which is in contrast with the dome shape of the doping
dependent SC gap parameter, and therefore should be verified by
further experiments.

\acknowledgments

This work was supported by the National Natural Science Foundation
of China under Grant No. 10774015, and the funds from the Ministry
of Science and Technology of China under Grant Nos. 2006CB601002 and
2006CB921300.


\begin{references}
\bibitem [*] {add} To whom correspondence should be addressed.
\end{references}
\end{document}